\documentclass[submission,Phys]{SciPost}
\usepackage[T1]{fontenc}
\usepackage{lmodern} 
\usepackage[bitstream-charter]{mathdesign}
\usepackage{microtype,bm,booktabs,placeins}
\hypersetup{colorlinks,
 linkcolor={red!50!black},
 citecolor={blue!50!black},
 urlcolor={blue!80!black},
}
\DeclareSymbolFont{usualmathcal}{OMS}{cmsy}{m}{n}
\DeclareSymbolFontAlphabet{\mathcal}{usualmathcal}
\newcommand{\ellB}{\ell_B}
\newcommand{\MBZ}{\mathrm{MBZ}}
\newcommand{\Tr}{\operatorname{Tr}}
\newcommand{\tr}{\operatorname{tr}}
\newcommand{\dd}{\mathrm{d}}
\newcommand{\Ec}{E_c}
\newcommand{\sh}{\mathsf{S}}
\newcommand{\hs}{\mathrm{HS}}

\begin{document}
\begin{center}
{\Large\bfseries Gap-label bounds and optical reconstruction\\[0.2em]
of Landau-subband quantum geometry\par}
\end{center}
\begin{center}
\"Ozg\"ur Kelek\c{c}i\textsuperscript{1$\star$}
\end{center}
\begin{center}
{\bfseries 1} Department of Basic Sciences, Faculty of Engineering,\\
University of Turkish Aeronautical Association,\\
Etimesgut/Ankara 06790, T\"urkiye\\[0.3em]
${}^\star$ {\small\sffamily\href{mailto:okelekci@thk.edu.tr}{okelekci@thk.edu.tr}}
\end{center}
\section*{Abstract}
{\bfseries

We derive a magnetic gap-label lower bound on the internal quantum weight of Landau subbands that remains nonzero even when the occupied Chern number vanishes. Using the second fundamental forms of nested subbundles to decompose the quantum geometric tensor, we isolate the constraint imposed by the uniform parent curvature and show that the excess quantum weight above the bound equals twice the weaker circular transition weight. For a noninteracting parabolic band with a periodic scalar potential, Landau-level mixing preserves the integrated relative curvature while the occupied direct gap remains open. Nevertheless, uncorrected sub-cyclotron absorption need not satisfy the projected bound. Using exact circular-dipole relations, we reconstruct the low-frequency guiding-centre weight and signed dichroism from circularly polarized absorption. This reconstruction removes the first-order orbit-polarization contribution. For an insulating filling of an isolated lowest-Landau-level-derived cluster, a potential-gradient estimate bounds the omitted remote-transition weight quadratically at weak coupling. This yields a finite-window lower bound on reconstructed quantum weight and bounds the deviation of reconstructed dichroism from the integrated relative curvature.
\par}

\noindent\rule{\textwidth}{0.8pt}
\tableofcontents
\noindent\rule{\textwidth}{0.8pt}

\section{Introduction}
The ordinary Chern number constrains the quantum geometry of an occupied band, but it does not exhaust the constraints inherited from a larger quantum-state space. A split Landau level makes this distinction concrete. Its occupied subbands lie in a parent bundle with uniform central curvature, so transitions within the level retain information about that embedding even when the occupied Chern number vanishes. The quantum geometric tensor (QGT), whose metric and curvature originate in the geometry of quantum states and adiabatic transport~\cite{ProvostVallee1980,Berry1984,Simon1983}, provides a natural language for this question. Its role in band dynamics and response~\cite{Xiao2010,Kolodrubetz2017,Yu2025} is particularly direct in insulators, where the metric controls the gauge-invariant Wannier spread, the localization tensor, and inverse-frequency-weighted absorption~\cite{MarzariVanderbilt1997,SWM,Resta2011,OnishiFu2025}. Positivity links these quantities to Chern numbers~\cite{Roy2014} and topological gap bounds~\cite{OnishiFu2024}, while saturation imposes a holomorphicity condition~\cite{MeraOzawa2021}.

In a magnetic field, cyclotron and guiding-centre motion separate the fixed geometry of a Landau orbit from the geometry changed by a periodic potential~\cite{GMP1986,Haldane2011,Parameswaran2013}. Magnetic Bloch theory and Hall gap labelling then connect the split spectrum to two integers~\cite{Hofstadter1976,TKNN,Streda,Dana}. The corresponding nested-bundle description decomposes the QGT into positive inner and outer tensors, expressed through second fundamental forms~\cite{OMP,Mitscherling2025}. Its mathematical foundation is the classical Gauss--Codazzi and Chern--Weil geometry of Hermitian subbundles, including the curvature identities used in Hermitian--Einstein stability theory~\cite{GriffithsHarris1978,Kobayashi1987,Lubke1983}. Recent sector bounds~\cite{Hung2026} and bounds in topologically trivial spinful systems~\cite{JSL2025} emphasize that geometric information can survive beyond the occupied Chern number. For a Landau parent, the additional information is particularly explicit. The internal curvature integrates to the Hall integer minus the occupied rank fraction. Our use of the classical identity is to identify this residual with a magnetic gap label and follow its physical response beyond projection.

That last step is essential for spectroscopy. Strict projection constrains projected position, whereas an electric field couples to the full position operator. Restricted optical sum rules address this distinction in correlated insulators~\cite{Mao2025}, and optical studies of moir\'e fractional Chern insulators include periodically modulated Landau levels as a theoretical setting~\cite{Paul2026}. Hall-absorption and dichroic sum rules likewise constrain how transverse response is distributed across frequency~\cite{ZH2026,SouzaVanderbilt2008}, and for a continuum Landau level with a periodic potential, partial Hall spectral weights provide an optical signature of Landau-level mixing~\cite{ZH2026}. These works motivate a precise question for a noninteracting parabolic band. Which low-frequency optical combination retains the relative-curvature constraint when Landau levels mix, and how much of that constraint can escape the measured window? An inequality for projected geometry alone does not answer this question because mixing changes both the low-energy projectors and the physical dipoles.

Circular spectroscopy supplies the required channel resolution. Integrated circular excitation rates can measure band topology~\cite{Tran2017,Asteria2019}, and circular dichroism has also been developed for confined quantum Hall systems, where the edge response probes the many-body Chern number and geometric anisotropy~\cite{UnalNardinGoldman2025}. Periodic-driving protocols also probe the quantum metric~\cite{OzawaGoldman2018}. For the continuum Hamiltonian studied here, the cyclotron and guiding-centre decomposition yields an elementary exact relation between their circular dipoles. We use this kinematic relation together with a curved guiding-centre connection to reconstruct both a metric weight and a signed dichroic response. The latter resolves the rational gap-label residual with an explicit error bound. This is distinct from identifying an unweighted circular response with a Chern number or assuming that a projected inequality remains valid for the physical sub-cyclotron spectrum.

The uniform curvature of the Landau parent allows us to translate the classical subbundle inequality into an internal quantum-weight bound determined by the magnetic gap label. In particular, this constraint remains nontrivial for a Chern-trivial occupied set, whose internal Wannier spread cannot be made arbitrarily small. The second fundamental form also resolves the excess above the bound into circular transition weights, with saturation expressed by a relative holomorphicity condition. We then examine how this geometry extends beyond strict projection. The guiding-centre connection retains the same integrated relative curvature when Landau levels mix, and circular optical moments reconstruct its low-frequency metric weight and signed dichroism. This reconstruction removes the generic first-order orbit contribution that contaminates uncorrected absorption. Finally, for an insulating filling of an isolated lowest-Landau-level-derived cluster, we control the missing remote transitions by a potential-gradient estimate that is quadratic in the potential amplitude at weak coupling. It gives both a two-sided error estimate for reconstructed dichroism and a finite-window lower bound for the reconstructed quantum weight. The numerical examples further show why the remote contribution remains necessary even when the low-frequency window is spectrally separated from higher Landau levels.

Section~\ref{sec:geometry} establishes the geometry and its localization consequences. Sections~\ref{sec:exact} and \ref{sec:remainder} carry the constraint into circular response and bound the omitted transitions. The examples in Sec.~\ref{sec:examples} then connect geometric excess, orbit polarization, and spectral transfer. Section~\ref{sec:discussion} addresses the physical regime and measurement requirements, followed by the conclusions in Sec.~\ref{sec:conclusions}.

\section{Geometry of Landau subbands}
\label{sec:geometry}
The magnetic subbands inherit geometry from the successive embeddings of the occupied space inside the parent Landau level and of the parent level inside the full Bloch Hilbert space. Keeping these embeddings distinct allows the fixed cyclotron contribution to be separated from the internal geometry that the periodic potential can change.

\subsection{Magnetic Bloch bundles and conventions}
We consider a two-dimensional, spin-polarized particle with charge $q_e$, $e=|q_e|$, in a uniform field with $q_eB>0$. Define
\begin{equation}
 \ellB^2=\frac{\hbar}{q_eB},\qquad \omega_c=\frac{q_eB}{m},\qquad \Ec=\hbar\omega_c.
 \label{eq:scales}
\end{equation}
The periodic scalar potential has $p/q$ flux quanta per primitive cell, with $p,q$ positive and coprime. A magnetic cell contains $q$ primitive cells and has area $A_{\rm mag}=2\pi p\ellB^2$. Thus $|\MBZ|=2\pi/(p\ellB^2)$, and a single Landau level defines a rank-$p$ bundle $E$ over the magnetic Brillouin torus.

Throughout, for a Hermitian QGT we set $g_{ij}=\operatorname{Re}Q_{ij}$ and $F=2\operatorname{Im}Q_{xy}$. For an anti-Hermitian connection curvature $\mathcal F$, the corresponding oriented Chern number is
\begin{equation}
 C=\frac{1}{2\pi i}\int_{\MBZ}\Tr\mathcal F_{xy}\,\dd^2k.
 \label{eq:chernconv}
\end{equation}
These conventions assign $C(E)=+1$ to a complete Landau level and $\sigma_{xy}=t_r e^2/h$, with $j_x=\sigma_{xy}E_y$. They reverse the sign of the curvature defined from the alternative Berry connection $i\langle u|\dd u\rangle$. The tensor formulas, Hall labels, and circular polarizations below consistently use this orientation.

The scalar $r$ below denotes an occupied rank, while bold $\bm r$ and its Cartesian or circular components denote position. Let $L\subset E$ be a smooth rank-$r$ subbundle, with orthogonal projectors $P_L\le P_E$ and oriented Chern number $t_r$. A direct band gap suffices to define this geometry. When $L$ is the occupied projector of a spectral insulator, the TKNN--St\v{r}eda labels satisfy~\cite{TKNN,Streda,Dana}
\begin{equation}
 r=qs_r+pt_r,\qquad \nu=\frac rp.
 \label{eq:gaplabels}
\end{equation}
The optical statements will additionally assume a global insulating gap and zero temperature.

\subsection{QGT decomposition and relative curvature}
The two embeddings can be expressed without choosing an occupied eigenvector frame. In the full magnetic Bloch Hilbert bundle, define
\begin{align}
 \sh_i^{\rm in}&=(P_E-P_L)(\partial_iP_L)P_L,\nonumber\\
 \sh_i^{\rm out}&=(1-P_E)(\partial_iP_L)P_L
                =(1-P_E)(\partial_iP_E)P_L .
 \label{eq:shape}
\end{align}
Here $\partial_i$ denotes the ambient Bloch derivative. In a moving magnetic basis it includes that basis's connection. The first map is the second fundamental form of $L$ inside $E$, equipped with the induced connection $\nabla^E=P_E\dd$. The second describes how $E$ itself sits in the full Hilbert space, restricted to $L$. Since their targets are orthogonal,
\begin{align}
 Q^L_{ij}&=\Tr\big[P_L(\partial_iP_L)(1-P_L)(\partial_jP_L)P_L\big]
             =Q^{\rm in}_{ij}+Q^{\rm out}_{ij},\nonumber\\
 Q^{\rm in}_{ij}&=\Tr[(\sh_i^{\rm in})^\dagger\sh_j^{\rm in}],\qquad
 Q^{\rm out}_{ij}=\Tr[(\sh_i^{\rm out})^\dagger\sh_j^{\rm out}].
 \label{eq:decomposition}
\end{align}
This is the subbundle QGT construction~\cite{OMP}, expressed in projector form~\cite{Mitscherling2025}. Each tensor is a Gram matrix of a second fundamental form, so positivity is preserved separately in the two transition sectors. The inner tensor is defined relative to the induced parent connection $\nabla^E$, including its curvature.

We use the Euclidean metric in physical reciprocal-space coordinates to define
\begin{equation}
 \Omega_I^{\rm gc}=\frac{A_{\rm mag}}{(2\pi)^2}\int_{\MBZ}\tr g^{\rm in}\,\dd^2k,
 \qquad K_{\rm gc}=\frac{\Omega_I^{\rm gc}}{p\ellB^2}
                 =\frac{1}{2\pi}\int_{\MBZ}\tr g^{\rm in}\,\dd^2k.
 \label{eq:weight}
\end{equation}
The trace $\Tr$ acts on quantum states, while $\tr g=g_{xx}+g_{yy}$ acts on coordinate indices. $\Omega_I^{\rm gc}$ has units of length squared and $K_{\rm gc}$ is dimensionless.

The curvature of the parent connection constrains how small this inner weight can be. Consider a Hermitian rank-$p$ bundle $E$ over the flat Brillouin torus whose unitary connection has central curvature, $\mathcal F^E=p^{-1}\Tr(\mathcal F^E)\,1_E$, and a rank-$r$ subbundle $L$ carrying the induced connection. The Gauss equation expresses the curvature of $L$ through that of $E$ and the second fundamental form. Traced over $L$, and with the central parent curvature contributing the fraction $r/p$ of its trace, it reads
\begin{equation}
 \Tr\mathcal F^L_{xy}
 =\frac rp\Tr\mathcal F^E_{xy}
  +Q^{\rm in}_{xy}-Q^{\rm in}_{yx}.
 \label{eq:gauss}
\end{equation}
The last two terms equal $iF^{\rm in}$. Integrating Eq.~\eqref{eq:gauss} with the convention of Eq.~\eqref{eq:chernconv} therefore relates the relative curvature to the two Chern numbers, $(2\pi)^{-1}\int F^{\rm in}=C(L)-(r/p)C(E)$. The inner tensor is a Gram matrix by Eq.~\eqref{eq:decomposition}, so $\tr g^{\rm in}\ge|F^{\rm in}|$ at every point of the MBZ. Integrating this pointwise inequality yields the classical Chern--Weil bound~\cite{GriffithsHarris1978,Kobayashi1987,Lubke1983}
\begin{equation}
 \frac{1}{2\pi}\int_{\MBZ}\tr g^{\rm in}\,\dd^2k
 \ \ge\ \left|C(L)-\frac rp C(E)\right|.
 \label{eq:generalbound}
\end{equation}
Its left-hand side is the inner quantum weight of Eq.~\eqref{eq:weight}, while its right-hand side involves only the ranks and Chern numbers of $L$ and $E$.

A uniform Landau level provides such a projectively flat parent. For its $n$th level, the oscillator matrix elements in Appendix~\ref{app:conventions} give the scalar parent QGT
\begin{equation}
 P_E(\partial_iP_E)(1-P_E)(\partial_jP_E)P_E
 =\frac{\ellB^2}{2}\big[(2n+1)\delta_{ij}+i\epsilon_{ij}\big]P_E,
 \qquad \epsilon_{xy}=1.
 \label{eq:parentQGT}
\end{equation}
Its induced curvature is therefore $\mathcal F^E_{xy}=i\ellB^2 1_E$, which makes the parent connection projectively flat. Equations~\eqref{eq:decomposition} and \eqref{eq:parentQGT} give
\begin{equation}
 \frac{1}{2\pi}\int F^{\rm out}=\nu,\qquad
 \frac{1}{2\pi}\int F^{\rm in}=t_r-\nu\equiv\kappa.
 \label{eq:relativecurvature}
\end{equation}
Here and below unqualified curvature integrals are over the MBZ. The relative contribution $F^{\rm in}$ need not integrate to an integer on this torus. Its rational value nevertheless has an ordinary bundle interpretation on a cover. Let $\pi\colon T'\to\MBZ$ be a $p$-fold covering with the pulled-back flat metric, and choose a line bundle $\Lambda$ with curvature $i\ellB^2\dd k_x\wedge\dd k_y$. Since $|T'|=2\pi/\ellB^2$, the line bundle $\Lambda$ has oriented Chern number one. The bundle $\pi^*E\otimes\Lambda^{-1}$ is flat, and the induced subbundle satisfies
\begin{equation}
 C(\pi^*L\otimes\Lambda^{-1})=pt_r-r=-qs_r.
 \label{eq:cover}
\end{equation}
Tensoring both bundles by the same line bundle leaves their second fundamental form unchanged. The usual metric--Chern bound on the cover~\cite{Roy2014,MeraOzawa2021}, followed by division by $p$, is therefore precisely the bound below. Flatness here does not require trivial holonomy. This construction explains the integer $p\kappa$ without assuming an equivalence to a particular dual Harper Hamiltonian.

For a subbundle of one Landau level, $C(E)=1$ and $C(L)=t_r$, so Eq.~\eqref{eq:generalbound} becomes the gap-label bound on the internal quantum weight,
\begin{equation}
 K_{\rm gc}\ge|\kappa|=\left|t_r-\frac rp\right|.
 \label{eq:projectedbound}
\end{equation}
For a spectral filling, Eq.~\eqref{eq:gaplabels} gives $|\kappa|=(q/p)|s_r|$, so the bound reads $K_{\rm gc}\ge(q/p)|s_r|$, or equivalently $\Omega_I^{\rm gc}\ge q|s_r|\ellB^2$.

Every proper subbundle, $0<r<p$, has $|t_r-r/p|\ge1/p$, since $pt_r-r$ is a nonzero integer. For a spectral gap, $s_r\ne0$ gives the stronger label-based minimum $q/p$. In particular, a Chern-trivial occupied set obeys $K_{\rm gc}\ge r/p$. Geometrically, a proper subbundle cannot be parallel under the parent connection because $\sh^{\rm in}=0$ would require the integer $t_r$ to equal the noninteger $r/p$. This obstruction concerns the quantum-state bundle over momentum space and requires no real-space curvature. The inner bound is identical for every Landau index, while the outer spread is $(2n+1)r\ellB^2$. Thus the total projected quantum weight $K=\Omega_I/(p\ellB^2)$ obeys
\begin{equation}
 K\ge(2n+1)\nu+|t_r-\nu|\ge|t_r|.
 \label{eq:totalprojected}
\end{equation}
In the lowest Landau level (LLL), the gauge-invariant spread satisfies
\begin{equation}
 \Omega_I=\Omega_I^{\rm gc}+r\ellB^2
 \ge(r+q|s_r|)\ellB^2.
 \label{eq:wannierbound}
\end{equation}
For a Chern-trivial sector admitting localized composite Wannier functions, the average gauge-invariant spread is therefore at least $2\ellB^2$ per function, twice the cyclotron contribution alone. Equation~\eqref{eq:totalprojected} exceeds the ordinary bound $|t_r|$ by $(2n+2)\nu$ whenever $t_r\le0$, and by $2n\nu$ for $t_r\ge1$ when $n\ge1$. Its right-hand side equals $|t_r|$ only for $n=0$ and integer $t_r\ge1$. The additional information is then the resolution into internal and cyclotron sectors.

\subsection{Saturation and circular transition weight}
The lower bound fixes a minimum amount of internal geometry. To characterize how closely a given subband approaches that minimum, the same second fundamental form identifies the excess above the bound. Directly from Eq.~\eqref{eq:decomposition},
\begin{align}
 \tr g^{\rm in}\mp F^{\rm in}
 &=\|\sh_x^{\rm in}\pm i\sh_y^{\rm in}\|_{\hs}^{2},\nonumber\\
 K_{\rm gc}\mp\kappa
 &=\frac{1}{2\pi}\int\|\sh_x^{\rm in}\pm i\sh_y^{\rm in}\|_{\hs}^{2}\,\dd^2k.
 \label{eq:squares}
\end{align}
Equality in Eq.~\eqref{eq:projectedbound} requires one fixed chirality of the second fundamental form to vanish everywhere. For $\kappa>0$ and $z=k_x+ik_y$, this is
\begin{equation}
 (P_E-P_L)(\nabla^E_{\bar z}P_L)P_L=0,
 \label{eq:holomorphic}
\end{equation}
where $\nabla^E_{\bar z}P_L$ is the covariant derivative of the endomorphism. It states that $L$ is a holomorphic subbundle of $(E,\bar\partial_E)$ with $\bar\partial_E=\nabla^E_{\bar z}$. For $\kappa<0$ the corresponding structure uses $z=k_x-ik_y$. This relative condition is the subbundle analogue of the trace and ideal-band conditions~\cite{MeraOzawa2021,Wang2021,Estienne2023}. It is closely related to the special trace-condition class of vortexable bands, rather than to every more general vortexability criterion~\cite{Ledwith2023}.

The integrated excess has two nonnegative sources,
\begin{equation}
 K_{\rm gc}-|\kappa|
 =\frac{1}{2\pi}\int(\tr g^{\rm in}-|F^{\rm in}|)\,\dd^2k
  +\left(\frac{1}{2\pi}\int|F^{\rm in}|\,\dd^2k-|\kappa|\right).
 \label{eq:excesssplit}
\end{equation}
The first measures local failure of the trace condition. The second measures cancellation between regions of opposite relative curvature.

For projected circular transitions, the two integrated weights are $(K_{\rm gc}\pm\kappa)/2$. Their weaker component and its fraction of the total are therefore
\begin{equation}
 K_{\rm weak}^{(0)}=\frac{K_{\rm gc}-|\kappa|}{2},\qquad
 f_{\rm weak}^{(0)}=\frac{K_{\rm gc}-|\kappa|}{2K_{\rm gc}}.
 \label{eq:weak}
\end{equation}
The geometric excess is twice the weaker circular weight. It therefore measures the departure from a single-chirality transition geometry directly.

For the complementary projector $P_E-P_L$, the internal tensor is the transpose of $Q^{\rm in}[P_L]$. The two metrics coincide and their relative curvatures have opposite signs. Consequently, the summed internal correction considered for complementary sectors by Hung \emph{et al.}~\cite{Hung2026} is $K_c=2K_{\rm gc}$ in this setting. For an LLL parent, whose total quantum weight is one, their general sector bound gives $K_c\ge |t_r|+|1-t_r|-1$, whereas Eq.~\eqref{eq:projectedbound} gives $K_c\ge2|t_r-\nu|$. The improvement is $2\nu$ for integer $t_r\le0$ and $2(1-\nu)$ for integer $t_r\ge1$, at least $2/p$ for a proper subbundle. Their notation $Q=g-i\Omega^{\rm H}/2$, with the superscript added here to distinguish their Berry curvature from a spread, translates as $\Omega^{\rm H}=-F$, with the comparison made in a common Hall orientation. The additional information is the signed relative curvature fixed by the uniform parent. Summing the two complementary tensors removes that information.

\section{Optical reconstruction beyond strict projection}
\label{sec:exact}
The projected result constrains transitions within a fixed parent Landau level. A physical optical experiment probes the full position operator, however, and its low-frequency matrix elements change when higher levels mix into the occupied states. The next step is to identify the optical combination that continues to represent guiding-centre geometry at finite cyclotron gap.

\subsection{Circular dipoles and optical moments}
Retaining all Landau levels, consider
\begin{equation}
 H=\frac{\bm\pi^2}{2m}+V(\bm r),\qquad
 \bm\pi=-i\hbar\nabla-q_e\bm A,\qquad V=\lambda v.
 \label{eq:H}
\end{equation}
The potential $v$ is fixed, smooth, real, bounded, and periodic. The amplitude $\lambda$ is dimensionless and $v$ has units of energy. The argument requires a parabolic kinetic energy and a scalar potential. Let $P$ be the rank-$r$ occupied projector below a global insulating gap. Define
\begin{equation}
 \bm\eta=\frac{\hat z\times\bm\pi}{q_eB},\qquad
 \bm R=\bm r-\bm\eta,\qquad
 r_\pm=\frac{r_x\mp ir_y}{\sqrt2},\quad
 R_\pm=\frac{R_x\mp iR_y}{\sqrt2}.
 \label{eq:coordinates}
\end{equation}
The separation of cyclotron and guiding-centre motion is standard magnetic kinematics and also underlies magnetic-Bloch semiclassics~\cite{ChangNiu1996}. We use its elementary dipole consequence as the starting point for the curvature and finite-window bounds, rather than as an independent geometric principle. For distinct exact eigenstates, $[H,r_i]=-i\hbar\pi_i/m$ gives $\bm\pi_{ba}=im\omega_{ba}\bm r_{ba}$, where $\omega_{ba}=(E_b-E_a)/\hbar$. Hence
\begin{equation}
 (R_\pm)_{ba}=\left(1\mp\frac{\omega_{ba}}{\omega_c}\right)(r_\pm)_{ba}.
 \label{eq:dipole}
\end{equation}
This kinematic identity is exact and remains regular at $\omega_{ba}=\omega_c$ in the displayed form. With time dependence $e^{-i\omega t}$, the plus channel is cyclotron-active for $q_eB>0$.

For the normalized complex polarizations $\bm e_\pm=(\hat x\mp i\hat y)/\sqrt2$ and time dependence $e^{-i\omega t}$, the absorptive conductivities are
\begin{equation}
 a_\pm=\operatorname{Re}(\bm e_\pm^\dagger\sigma\bm e_\pm)
 =\tfrac12\operatorname{Re}(\sigma_{xx}+\sigma_{yy})
 \pm\tfrac12\operatorname{Im}(\sigma_{xy}-\sigma_{yx})\ge0.
 \label{eq:circularconductivity}
\end{equation}
This formula also applies to an anisotropic conductivity. In our $q_eB>0$ orientation, the plus field rotates from $+\hat x$ toward $-\hat y$. For electrons in a laboratory field $B\hat z$ with $B>0$, $q_eB<0$ and the channel labels must be interchanged when converting to the laboratory axes. The cyclotron-active field then has complex polarization $(\hat x+i\hat y)/\sqrt2$ and rotates from $+\hat x$ toward $+\hat y$. For an angular-frequency cutoff $\omega_*>0$, define
\begin{equation}
 M_j^\Sigma=\int_0^{\omega_*}\omega^j(a_++a_-)\,\dd\omega,\qquad
 M_j^D=\int_0^{\omega_*}\omega^j(a_+-a_-)\,\dd\omega.
 \label{eq:moments}
\end{equation}
The low-frequency physical quantum weight is $K_{\rm sub}=(2\hbar/e^2)M_{-1}^\Sigma$~\cite{SWM,OnishiFu2025}. In strict projection, when the window contains all transitions from $L$ to $E\setminus L$, the projected optical sum rule gives $K_{\rm sub}^{(0)}=K_{\rm gc}$~\cite{Mao2025,Hung2026}. At finite gap, Eq.~\eqref{eq:dipole} instead gives the exact reconstruction
\begin{equation}
 K_R^{\rm low}=\frac{2\hbar}{e^2}
 \left[M_{-1}^\Sigma-\frac{2M_0^D}{\omega_c}
                         +\frac{M_1^\Sigma}{\omega_c^2}\right].
 \label{eq:reconstruction}
\end{equation}
The same channel weights reconstruct the signed dichroic combination,
\begin{equation}
 D_R^{\rm low}=\frac{2\hbar}{e^2}
 \left[M_{-1}^D-\frac{2M_0^\Sigma}{\omega_c}
                    +\frac{M_1^D}{\omega_c^2}\right].
 \label{eq:dichroism}
\end{equation}
Each circular channel is weighted by $(1\mp\omega/\omega_c)^2$ in its negative-first moment. In particular, the leading correction uses the \emph{zeroth} dichroic conductivity moment. $K_{\rm sub}$ alone need not satisfy the projected bound.

\subsection{Relative curvature of the mixed bands}
Equation~\eqref{eq:reconstruction} identifies a guiding-centre transition weight in the physical spectrum. To determine whether this weight retains a geometric lower bound, we must establish which curvature accompanies it. The appropriate QGT is defined through a connection on the magnetic Bloch fibres,
\begin{equation}
 D_i=\partial_{k_i}+i\eta_i(\bm k),\qquad
 D_iP:=[D_i,P],\qquad
 \sh_i^R=(1-P)(D_iP)P.
 \label{eq:D}
\end{equation}
The notation $D_iP$ denotes a covariant derivative of an endomorphism, not composition of differential operators. Between occupied and empty states, $\sh_i^R=-i(1-P)R_iP$. We define
\begin{equation}
 Q^R_{ij}=\Tr[(\sh_i^R)^\dagger\sh_j^R],\qquad
 F_R=2\operatorname{Im}Q^R_{xy},\qquad
 K_R=\frac{1}{2\pi}\int\tr g^R\,\dd^2k.
 \label{eq:QR}
\end{equation}
In this formulation, the transition expression $\Tr[PR_i(1-P)R_jP]$ is understood through the fibrewise off-diagonal blocks in Eq.~\eqref{eq:D}. These blocks define the second fundamental form even though the full position operator acts differentially in momentum space.

This tensor inherits the integrated constraint of strict projection, because the connection in Eq.~\eqref{eq:D} carries the same central curvature as the parent connection of a single Landau level, $[D_x,D_y]=i\ellB^2$ (Appendix~\ref{app:curvature}). Let $P$ be a smooth finite-rank isolated spectral projector of Eq.~\eqref{eq:H}, with rank $r$ and oriented Chern number $t_r$. The traced Gauss equation for the induced connection $PDP$ then reads $\Tr\mathcal F^{PDP}_{xy}=i\ellB^2r+2i\operatorname{Im}Q^R_{xy}$. The connections $PDP$ and $P\dd P$ differ by the endomorphism-valued one-form $iP\eta P$, so their traced curvatures differ by an exact two-form and both integrate to the occupied Chern number $t_r$. Integrating over the MBZ with $|\MBZ|=2\pi/(p\ellB^2)$ and dividing by $2\pi i$ therefore gives
\begin{equation}
 \frac{1}{2\pi}\int_{\MBZ}F_R\,\dd^2k=t_r-\frac rp=\kappa.
 \label{eq:exactcurv}
\end{equation}
Appendix~\ref{app:curvature} supplies the sewing and domain details needed for this argument. The identity holds at any coupling at which the projector remains isolated, and it requires only a direct gap. Landau-level mixing can therefore reshape the guiding-centre metric and the local distribution of $F_R$, but not its integral.

The curvature identity now connects the reconstructed spectrum to the projected bound. For absorption out of a zero-temperature insulating ground state, split the positive transition tensor into $0<\omega_{ba}\le\omega_*$ and $\omega_{ba}>\omega_*$ sectors. A line exactly at the cutoff is included in the low sector in all moments and excluded from the remote integral. Write their integrated weights and curvatures as $K_R^{\rm low,rem}$ and $\kappa_{\rm low,rem}$. Positivity holds separately even if the cutoff does not define a common empty-band projector.

For a clean, noninteracting system at zero temperature with a global insulating gap, take $P$ to be the occupied projector of Eq.~\eqref{eq:H}. In each transition, Eq.~\eqref{eq:dipole} converts the circular strengths $|(r_\pm)_{ba}|^2$, reweighted by $(1\mp\omega_{ba}/\omega_c)^2$, into the guiding-centre strengths $|(R_\pm)_{ba}|^2$. Their signed difference is the contribution of that transition to $F_R$, so the reconstructed dichroism of Eq.~\eqref{eq:dichroism} measures the low-window relative curvature, $D_R^{\rm low}=\kappa_{\rm low}$. Equation~\eqref{eq:exactcurv} fixes $\kappa_{\rm low}+\kappa_{\rm rem}=\kappa$, and positivity of the remote tensor gives $|\kappa_{\rm rem}|\le K_R^{\rm rem}$. Hence
\begin{equation}
 D_R^{\rm low}=\kappa_{\rm low},\qquad
 |D_R^{\rm low}-\kappa|\le K_R^{\rm rem}.
 \label{eq:dichroicbound}
\end{equation}
Positivity of the low tensor gives in the same way $K_R^{\rm low}\ge|\kappa_{\rm low}|=|\kappa-\kappa_{\rm rem}|$. With the triangle inequality, this yields the metric counterpart of Eq.~\eqref{eq:dichroicbound}, in which Eq.~\eqref{eq:dipole} expresses the remote weight through the absorption above the cutoff,
\begin{align}
 K_R^{\rm low}&\ge |\kappa|-K_R^{\rm rem},\label{eq:finitebound}\\
 K_R^{\rm rem}&=\frac{2\hbar}{e^2}\int_{\omega>\omega_*}
 \frac{(1-\omega/\omega_c)^2a_++(1+\omega/\omega_c)^2a_-}{\omega}\,\dd\omega.
 \label{eq:remotemoment}
\end{align}

With all frequencies included, the statements become $D_R^{\rm tot}=\kappa$ and $K_R^{\rm tot}\ge|\kappa|$ at any coupling for which the occupied bundle stays isolated. A direct gap suffices for this geometric identity, whereas an absorption measurement with the occupied-to-empty interpretation requires the global gap and occupation assumptions stated above for the finite-window response.

The all-frequency dichroic identity also follows from three established sum rules, which provides an independent check of Eq.~\eqref{eq:exactcurv}. For $\omega_*\to\infty$ in Eq.~\eqref{eq:dichroism}, the Kramers--Kronig relation for the static Hall conductivity, the $f$-sum rule of the parabolic band, and the first-moment Hall-absorption sum rule~\cite{ZH2026}, a consequence of $[\pi_x,\pi_y]=i\hbar q_eB$, give $(2\hbar/e^2)M_{-1}^D=t_r$, $(2\hbar/e^2)\,2M_0^\Sigma/\omega_c=2\nu$, and $(2\hbar/e^2)M_1^D/\omega_c^2=\nu$, respectively. Hence $D_R^{\rm tot}=t_r-2\nu+\nu=\kappa$. The geometric construction identifies the reweighted transition strengths with a guiding-centre tensor that reduces to $Q^{\rm in}$ in strict projection. Its integrated-curvature identity requires only a direct gap. For an insulating filling, the low- and remote-transition decomposition and the potential-gradient estimate give the controlled finite-window result.

In strict projection, $D_{\rm sub}=(2\hbar/e^2)M_{-1}^D=\kappa$. This is the intra-level part of the Hall decomposition in Appendix~\ref{app:conventions}, with the parent contribution $\nu$ supplying the remaining Hall weight. The relation belongs to the same family of spectral constraints as quantized circular dichroism~\cite{Tran2017,Asteria2019}, but its finite-gap content is different. Equation~\eqref{eq:dichroism} removes the generic first-order term $2(2\hbar/e^2)M_0^\Sigma/\omega_c$ from $D_{\rm sub}$ and leaves only a remote-curvature error. The bound on that error is obtained next.

When an isolated LLL-derived cluster with projector $P_{\rm low}\ge P$ exists and the cutoff separates that cluster from all higher levels, the frequency decomposition has the direct subbundle interpretation
\begin{align}
 \sh_i^{R,\rm low}&=(P_{\rm low}-P)(D_iP)P,\nonumber\\
 \sh_i^{R,\rm rem}&=(1-P_{\rm low})(D_iP)P,\nonumber\\
 Q^R_{ij}&=\Tr[(\sh_i^{R,\rm low})^\dagger\sh_j^{R,\rm low}]
          +\Tr[(\sh_i^{R,\rm rem})^\dagger\sh_j^{R,\rm rem}].
 \label{eq:finiteshape}
\end{align}
The remote remainder is the integrated squared norm of this second, extrinsic component. The full tensor also obeys
\begin{equation}
 K_R^{\rm low}+K_R^{\rm rem}\mp\kappa
   =\frac{1}{2\pi}\int\|\sh_x^R\pm i\sh_y^R\|_{\hs}^2\,\dd^2k.
 \label{eq:finitesquares}
\end{equation}
Equations~\eqref{eq:squares} and \eqref{eq:finitesquares} express the projected and finite-gap constraints through the same geometric construction. Their difference is the transfer of a nonnegative part of the guiding-centre weight to transitions outside the chosen window. Bounding this part will turn the reconstruction into a constraint on low-frequency observables alone.

\section{A controlled bound for a finite optical window}
\label{sec:remainder}
The finite-window inequality is exact, but its remote term still involves transitions outside the measured frequency range. For an isolated lowest-Landau-level-derived cluster, that term can be bounded using only the strength and gradient of the scalar potential. The explicit optical estimates in this section are restricted to occupied subbands in the lowest cluster, although Sec.~\ref{sec:geometry} applies to any parent Landau level. The essential input is that the guiding centre commutes with the kinetic energy, so the potential controls all of its remote transition matrix elements.

Let $\delta=\tfrac12(\max V-\min V)$, so a constant shift $c$ makes $\|V-c\|_\infty=\delta$. The spectrum of $H-c$ lies within $\delta$ of the free Landau energies. For $\delta<\Ec/2$, the LLL-derived cluster is isolated from the next one by at least $\Ec-2\delta$.

Take $\delta<\Ec/2$ and an occupied rank-$r$ spectral projector $P$ contained in the LLL-derived cluster. Since the guiding centre commutes with the kinetic Hamiltonian, its commutators with $H$ come entirely from the potential, $[R_x,V]=-i\ellB^2\partial_yV$ and $[R_y,V]=i\ellB^2\partial_xV$. An occupied eigenstate $a$ and a remote eigenstate $b$ outside the cluster therefore obey
\begin{equation}
 (E_b-E_a)(R_i)_{ba}=[H,R_i]_{ba},\qquad E_b-E_a\ge\Ec-2\delta.
 \label{eq:commutatorbound}
\end{equation}
Dividing by the energy separation and applying Bessel's inequality bounds the total remote matrix-element weight of each occupied state,
\begin{equation}
 \sum_{b\notin\mathrm{low}}|(R_x)_{ba}|^2
 \le\frac{\|[H,R_x]|a\rangle\|^2}{(\Ec-2\delta)^2}
 \le\frac{\ellB^4\|\partial_yV\|_\infty^2}{(\Ec-2\delta)^2},
 \label{eq:bessel}
\end{equation}
and similarly with $x,y$ interchanged. Summing over the $r$ occupied states, integrating with $(2\pi)^{-1}\int_{\MBZ}$, and using $|\MBZ|=2\pi/(p\ellB^2)$, we find that the remainder to all states outside the cluster obeys
\begin{equation}
 K_R^{\rm rem}\le B_V,\qquad
 B_V=\frac rp\,
 \frac{\ellB^2\big(\|\partial_xV\|_\infty^2+\|\partial_yV\|_\infty^2\big)}{(\Ec-2\delta)^2}.
 \label{eq:gradientbound}
\end{equation}

Keeping the occupied expectation in the first step of Eq.~\eqref{eq:bessel} gives the sharper estimate
\begin{equation}
 K_R^{\rm rem}\le B_{\rm occ}\le B_V,\qquad
 B_{\rm occ}=\frac{\ellB^4}{2\pi(\Ec-2\delta)^2}
 \int_{\MBZ}\Tr\!\left[P\big((\partial_xV)^2+(\partial_yV)^2\big)\right]\dd^2k.
 \label{eq:occupiedbound}
\end{equation}
Unlike $B_V$, this refinement needs the occupied states. Either estimate may use an independently established lower bound on the occupied-to-remote separation in place of $\Ec-2\delta$. A sampled band minimum alone is not a rigorous enclosure. The bound is invariant under adding a constant to $V$ and is $O(\lambda^2)$ for fixed $v$ and $\Ec$. The same order follows because the remote second fundamental form vanishes for the unmixed LLL and starts linearly in $\lambda$.

We now combine the remote-transition estimate with the optical reconstruction. Assume a global insulating gap within the LLL-derived cluster and $\delta<\Ec/4$. All transitions within the lowest cluster then have energy at most $2\delta$, whereas remote transitions have energy at least $\Ec-2\delta$. Any cutoff strictly inside
\begin{equation}
 \frac{2\delta}{\hbar}<\omega_*<\frac{\Ec-2\delta}{\hbar}
 \label{eq:window}
\end{equation}
therefore selects precisely the two sectors of Eq.~\eqref{eq:finiteshape}, and Eqs.~\eqref{eq:dichroicbound}, \eqref{eq:finitebound}, and \eqref{eq:gradientbound} combine into the three-moment results
\begin{equation}
 |D_R^{\rm low}-(t_r-r/p)|\le B_V,\qquad
 K_R^{\rm low}\ge|t_r-r/p|-B_V.
 \label{eq:mainoptical}
\end{equation}
In particular, $-(p/q)D_R^{\rm low}$ estimates the integer $s_r$ with absolute error at most $(p/q)B_V$. This is controlled proximity to a rational response, not exact finite-window quantization. A form without the first moment follows from positivity of $a_\pm$, which gives $M_1^\Sigma\le\omega_*^2M_{-1}^\Sigma$. Substituting this upper bound into Eq.~\eqref{eq:reconstruction} and using Eq.~\eqref{eq:mainoptical} gives the two-moment bound
\begin{equation}
 \frac{2\hbar}{e^2}\left[
  \left(1+\frac{\omega_*^2}{\omega_c^2}\right)M_{-1}^\Sigma
  -\frac{2M_0^D}{\omega_c}\right]
 \ge \left|t_r-\frac rp\right|-B_V.
 \label{eq:measurable}
\end{equation}

Equation~\eqref{eq:mainoptical} is the primary finite-window statement. The two-moment form trades the positive first moment for a cutoff-dependent upper bound. If a known upper edge $\omega_{\max}\le\omega_*$ contains the entire low-frequency absorption, $\omega_*$ in Eq.~\eqref{eq:measurable} may be replaced by $\omega_{\max}$ because $M_1^\Sigma\le\omega_{\max}^2M_{-1}^\Sigma$. A measured edge must include its uncertainty and any unresolved tail. The spectral enclosure in Eq.~\eqref{eq:window} remains an a priori way to choose the integration window.

The slack in Eq.~\eqref{eq:measurable} separates exactly into
\begin{equation}
 \frac{2\hbar}{e^2\omega_c^2}
   (\omega_*^2M_{-1}^\Sigma-M_1^\Sigma)
 +(B_V-K_R^{\rm rem})+(K_R^{\rm tot}-|\kappa|).
 \label{eq:slack}
\end{equation}
The three nonnegative terms are the moment surrogate, the overestimate of the remote weight, and the actual geometric excess. The metric right-hand side is positive when $B_V<|\kappa|$. The dichroic bound remains an absolute error estimate outside that range, although it may then give little information about the gap label.

Strict projection is recovered by taking $m\to0$ at fixed $B$, potential, and lattice cell, with an open projected insulating gap. Then $\Ec\to\infty$ while $\ellB$, $p/q$, and the projected geometry remain fixed. The reweighting tends to unity and $B_V\to0$, so Eq.~\eqref{eq:finitebound} reduces to Eq.~\eqref{eq:projectedbound}. This limit is distinct from changing the field and hence changing the magnetic geometry. Experimentally, the relevant mixing parameter is the potential scale divided by $\hbar\omega_c$, controlled through the field or the gate or lattice depth, rather than by changing the mass.

\section{Geometric and optical response of split Landau levels}
\label{sec:examples}
The preceding results distinguish the geometric excess in strict projection from the physical low-frequency weight at finite gap and the guiding-centre weight reconstructed from it. Two simple potentials illustrate how these quantities are related and why the optical correction matters even when Landau-level mixing is weak.

\subsection{Projected geometry and proximity to saturation}
For numerical results we set $\hbar=\omega_c=\ellB=e=1$. Potentials are expressed in units of $\Ec$, and the multiplier $\lambda$ in Eq.~\eqref{eq:H} remains dimensionless. We use square primitive cells with $q=1$, side $L=\sqrt{2\pi p}$, and $G=2\pi/L$. The two potentials and their gap labels are
\begin{align}
 \text{A}\quad &p=2,\qquad v_A=2\cos Gx+2\cos Gy-4\cos Gx\cos Gy,
                  \qquad (r,t_r,s_r)=(1,0,1),\label{eq:modelA}\\
 \text{B}\quad &p=3,\qquad v_B=2\cos Gx+2\cos Gy,
                  \qquad (r,t_r,s_r)=(2,1,-1).\label{eq:modelB}
\end{align}
Both are spectral insulators in the projected problem. Their Chern numbers are obtained from Wilson loops independently of the relative-curvature integrals. The continuum Hamiltonian is represented in a magnetic Bloch basis with Landau indices $0\le n\le24$, and the MBZ integrals use a $24\times24$ midpoint quadrature. Oscillator Fourier matrix elements are evaluated by Hermite-function quadrature, while momentum derivatives are taken covariantly with the full magnetic-basis connection, including the Landau-ladder terms. Off-diagonal derivatives are obtained from covariant Hamiltonian derivatives and interband energy differences. Table~\ref{tab:models} collects the projected geometry and the first-order response coefficients. The projected bounds are $|\kappa|=1/2$ for A and $1/3$ for B. The global gap of $\bar v$ is the difference between the empty-band minimum and the occupied-band maximum, evaluated on a $97\times97$ grid that includes the zone edges. For B it agrees with the closed form $(3-\sqrt3)e^{-\pi/6}$. A globally occupied insulating sector requires this indirect gap to be positive, whereas optical transitions are controlled by the minimum direct gap. The mean gap is the MBZ average of the adjacent-band gap $\varepsilon_r-\varepsilon_{r-1}$ of $\bar v$ and is not an average over all optical transitions. The last four rows give the first-order derivatives of $K_{\rm sub}$ with respect to $\lambda$ and the absolute mismatch between their sum and the continuum derivative.

\begin{table}[htbp]
\centering
\caption{Projected geometry and first-order slopes of $K_{\rm sub}$ for models A and B.}
\label{tab:models}
\begin{tabular}{lrr}
\toprule
 & A, $p=2$, $r=1$ & B, $p=3$, $r=2$\\
\midrule
Wilson-loop $t_r$ & 0 & 1 \\
$\kappa=t_r-r/p$ & $-1/2$ & $+1/3$ \\
$K_{\rm gc}$ & 0.50322367 & 0.40598213 \\
$K_{\rm gc}-|\kappa|$ & 0.00322367 & 0.07264879 \\
Weaker circular fraction (\%) & 0.3203 & 8.9473 \\
Global gap of $\bar v$ & 1.663037 & 0.751114 \\
Mean adjacent-band gap of $\bar v$ & 1.989028 & 1.439817 \\
Orbit polarization & $-1.854199$ & 0.756473 \\
Projector rotation & $-0.043370$ & 0.670627 \\
Sum of first-order terms & $-1.897569$ & 1.427100 \\
Absolute derivative mismatch & $<5\times10^{-9}$ & $<2\times10^{-8}$ \\
\bottomrule
\end{tabular}
\end{table}

Model A is close to saturating the bound with $K_{\rm gc}-|\kappa|=0.00322367$, so only $0.3203\%$ of its projected circular weight lies in the weaker channel. In model B the excess is $0.07264879$, corresponding to $8.9473\%$. These numbers quantify departure from the chirality condition in Eq.~\eqref{eq:squares}. A Chern-trivial occupied sector can be much closer to saturation than a Chern-nontrivial one. The pointwise square identities are verified to $4\times10^{-15}$ on the numerical grids, and their integrated values agree with the curvature-based excess to $5\times10^{-13}$. Both examples have single-signed $F^{\rm in}$ on the checked grids and in the extremum searches. Thus their excess is entirely the local trace-condition contribution in Eq.~\eqref{eq:excesssplit} within numerical accuracy. They do not test cancellation from sign-changing curvature or the dependence on $q>1$. Their total LLL gauge-invariant spreads are $2.00645\ellB^2$ and $3.21795\ellB^2$, against the respective lower bounds $2\ellB^2$ and $3\ellB^2$.

\subsection{Orbit polarization and projector rotation}
The small geometric excess of A makes the first correction to its physical optical weight particularly consequential. To identify that correction, we use a Schrieffer--Wolff separation of the lowest cluster~\cite{Bravyi2011}. At first order in $\lambda$, the physical low-frequency weight changes for two reasons. Virtual transitions to higher Landau levels polarize the cyclotron orbit, changing the physical dipole relative to $\bm R$. They also rotate the low-energy eigenprojectors through the second-order effective Hamiltonian
\begin{equation}
 H_{\rm eff}=\frac{\Ec}{2}+\lambda\bar v-\frac{\lambda^2}{\Ec}w+O(\lambda^3),
 \qquad \bar v=P_0vP_0,\quad
 w=\sum_{n\ge1}\frac{P_0vP_nvP_0}{n}.
 \label{eq:Heff}
\end{equation}
Although the Hamiltonian correction is second order, the eigenprojectors of its split LLL rotate at first order because the intra-level gaps themselves are $O(\lambda)$. Appendix~\ref{app:sw} derives the corresponding orbit and projector contributions. The orbit term is evaluated directly from Eq.~\eqref{eq:orbitslope}, and the projector term from a symmetric high-order difference of the effective projected operator. For comparison, the continuum derivative in Table~\ref{tab:models} uses Richardson-extrapolated differences at small positive and negative potential amplitudes. At negative amplitude, the order of the projected eigenvalues is reversed. Occupying $p-r$ states then selects the complement of the branch continued from the positive-amplitude rank-$r$ sector. Equality of complementary internal metrics, established in Sec.~\ref{sec:geometry}, makes this the appropriate continuation for the derivative.

In $K$ normalization, the orbit and projector contributions are $-1.854199$ and $-0.043370$ in A, and $0.756473$ and $0.670627$ in B. Their sums agree with continuum finite-difference derivatives to within $2\times10^{-8}$. Projector rotation is approximately $2.3\%$ of orbit polarization in A and $89\%$ in B, so it is not generically negligible. The exact optical reweighting removes the orbit contribution, while retaining the geometry of the rotated projectors. Their integrated relative curvature is unchanged as long as the gap stays open. The sign of the orbit term depends on energy-weighted transition curvatures and is not determined by the gap labels alone.

The same separation occurs in the Chern-trivial $p=2$ potential of Krishna, Ippoliti, and Bhatt~\cite{KIB2019}, which contains more Fourier harmonics. Using its published rounded coefficients, a Landau cutoff $n_{\max}=45$, and a $12\times12$ grid gives $K_{\rm gc}=0.50313833$. The orbit and projector slopes are $-8.74271145$ and $1.20228971$, summing to $-7.54042173$. The continuum extrapolation gives $-7.54042112$, a difference of about $6\times10^{-7}$. In this case projector rotation partly offsets orbit polarization, reinforcing that the two mechanisms need not have the same sign.

\subsection{Finite-gap absorption and the remote contribution}
\begin{figure}[t]
\centering
\includegraphics[width=\textwidth]{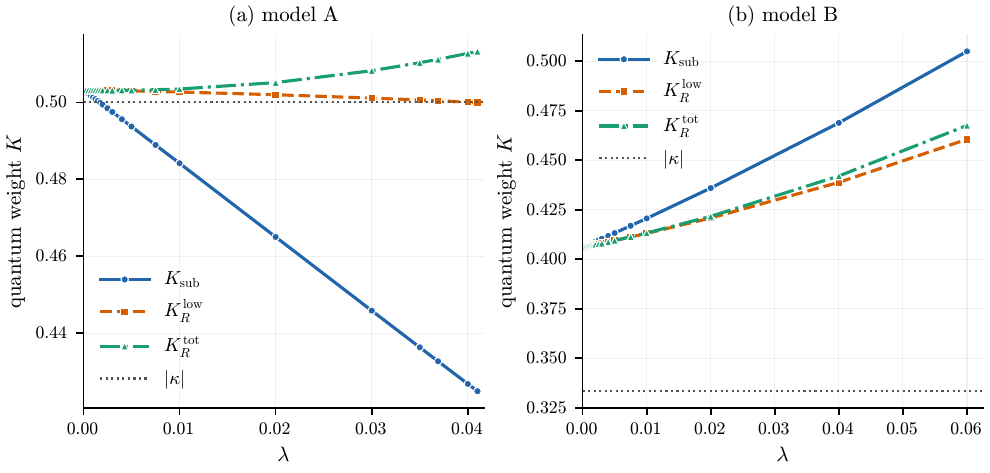}
\caption{Physical weight $K_{\rm sub}$ and reconstructed weights $K_R^{\rm low}$ and $K_R^{\rm tot}=K_R^{\rm low}+K_R^{\rm rem}$ for A and B. The dotted line is $|\kappa|$. Points at $\lambda=0$ denote the projected $\lambda\to0^+$ limit.}
\label{fig:spreads}
\end{figure}

The perturbative separation explains the initial slopes, while the exact reconstruction controls their continuation to finite potential strength. Figure~\ref{fig:spreads} compares $K_{\rm sub}$, $K_R^{\rm low}$, and $K_R^{\rm tot}$ over nearly the full a priori window range. Its zero-amplitude points are projected limits, since the unsplit Hamiltonian has no internal insulating gap. In A, the uncorrected physical weight crosses $|\kappa|=1/2$ at $\lambda_*=1.69736\times10^{-3}$, where the mean adjacent-band gap is $\bar\Delta_*/\Ec=3.39143\times10^{-3}$. Its small initial geometric excess and negative orbit slope explain why the projected inequality fails early for the physical spectrum. At $\lambda=0.01$, $K_{\rm sub}=0.484164$, whereas the reconstruction gives $K_R^{\rm low}=0.502700$.

The reconstructed low-window weight is itself not protected by the remainder-free bound. In A it crosses $1/2$ at $\lambda\simeq0.040782$, where $\delta/\Ec\simeq0.244692<1/4$. At $\lambda=0.041$, $K_R^{\rm low}=0.499976$ and $K_R^{\rm rem}=0.013301$, so $K_R^{\rm tot}=0.513276$ remains above $1/2$. This crossing occurs within the guaranteed optical-separation regime and demonstrates that the remote term is physically necessary. It is not visible on the scale of Fig.~\ref{fig:spreads}(a), because $K_R^{\rm low}-1/2=-2.4\times10^{-5}$ at $\lambda=0.041$. In B, both first-order contributions are positive and the reconstructed low-frequency weight rises throughout the displayed range.

The potential constants needed for the rigorous window and remainder estimates are analytic and are listed in Table~\ref{tab:constants}. Its last two columns give the amplitude range of the guaranteed window $\delta<\Ec/4$ and the range in which $B_V<|\kappa|$. For B the second range extends beyond the window, so the window condition is the binding one. For example, $\partial_xv_A=G\sin(Gx)[-2+4\cos(Gy)]$ has supremum $6G$. Writing $v_A$ as a bilinear function of the two cosines gives its oscillation $12$. With $\omega_*=1.01(2\delta/\hbar)$, the displayed finite-amplitude points through $\lambda=0.041$ in A and $0.06$ in B lie within Eq.~\eqref{eq:window}. Table~\ref{tab:scan} reports the reconstructed weights and both sides of the two-moment inequality on this extended range.

\begin{table}[htbp]
\centering
\caption{Analytic potential constants and amplitude ranges.}
\label{tab:constants}
\begin{tabular}{ccccc}
\toprule
Model & $\|\partial_xv\|_\infty=\|\partial_yv\|_\infty$ & $\operatorname{osc}v$ & $\delta<\Ec/4$ & $B_V<|\kappa|$ \\
\midrule
A & $6G$ & 12 & $\lambda<1/24$ & $\lambda<0.03698$ \\
B & $2G$ & 8 & $\lambda<1/16$ & $\lambda<0.07252$ \\
\bottomrule
\end{tabular}
\end{table}

\begin{table}[htbp]
\centering
\caption{Quantum weights and the two sides of Eq.~\eqref{eq:measurable} for $\omega_*=1.01(2\delta/\hbar)$.}
\label{tab:scan}
\begin{tabular}{crrrrrr}
\toprule
Model & $\lambda$ & $K_{\rm sub}$ & $K_R^{\rm low}$ & $K_R^{\rm tot}$ & LHS & RHS \\
\midrule
A & 0.001 & 0.501325 & 0.503179 & 0.503187 & 0.503251 & 0.499884 \\
A & 0.01 & 0.484164 & 0.502700 & 0.503487 & 0.509634 & 0.485395 \\
A & 0.02 & 0.465000 & 0.501997 & 0.505159 & 0.528598 & 0.421678 \\
A & 0.0369 & 0.432665 & 0.500423 & 0.511204 & 0.584458 & 0.003999 \\
A & 0.04 & 0.426773 & 0.500087 & 0.512749 & 0.597441 & $-0.169215$ \\
A & 0.041 & 0.424877 & 0.499976 & 0.513276 & 0.601788 & $-0.236703$ \\
B & 0.001 & 0.407413 & 0.406656 & 0.406658 & 0.406682 & 0.333322 \\
B & 0.01 & 0.420603 & 0.413034 & 0.413248 & 0.415710 & 0.332014 \\
B & 0.04 & 0.468862 & 0.438812 & 0.442038 & 0.486659 & 0.294682 \\
B & 0.06 & 0.504977 & 0.460559 & 0.467539 & 0.576705 & 0.184619 \\
\bottomrule
\end{tabular}
\end{table}

The uniform metric bound becomes uninformative in A at $\lambda\simeq0.03698$, before the separation condition fails at $1/24$. In B it remains positive throughout $\lambda<1/16$. This loss of a useful uniform estimate is distinct from the later crossing of $K_R^{\rm low}$ below its projected minimum in A. Both effects are consistent with the exact all-frequency constraint.

The signed response makes the benefit of reconstruction especially clear. Table~\ref{tab:dichroism} compares the uncorrected dichroism $D_{\rm sub}=(2\hbar/e^2)M_{-1}^D$ with $D_R^{\rm low}$. At $\lambda=0.01$, reconstruction reduces the deviation from $\kappa$ by factors of approximately 167 in A and 321 in B. The uncorrected deviation is first order in $\lambda$, whereas the reconstructed residual is second order. In these examples the remote tensor has little net chirality, so $|D_R^{\rm low}-\kappa|=|\kappa_{\rm rem}|$ is substantially smaller than its total weight $K_R^{\rm rem}$. No such cancellation is assumed in the rigorous bound.

\begin{table}[htbp]
\centering
\caption{Uncorrected and reconstructed dichroic deviations from $\kappa$ and the remote weight.}
\label{tab:dichroism}
\begin{tabular}{ccrrr}
\toprule
Model & $\lambda$ & $D_{\rm sub}-\kappa$ & $D_R^{\rm low}-\kappa$ & $K_R^{\rm rem}$ \\
\midrule
A & 0.001 & $1.868\times10^{-3}$ & $1.123\times10^{-6}$ & $7.826\times10^{-6}$ \\
A & 0.01 & $1.877\times10^{-2}$ & $1.121\times10^{-4}$ & $7.871\times10^{-4}$ \\
A & 0.04 & $7.542\times10^{-2}$ & $1.733\times10^{-3}$ & $1.266\times10^{-2}$ \\
B & 0.001 & $1.007\times10^{-3}$ & $3.294\times10^{-7}$ & $2.169\times10^{-6}$ \\
B & 0.01 & $1.021\times10^{-2}$ & $3.182\times10^{-5}$ & $2.134\times10^{-4}$ \\
B & 0.04 & $4.265\times10^{-2}$ & $4.432\times10^{-4}$ & $3.225\times10^{-3}$ \\
\bottomrule
\end{tabular}
\end{table}

\begin{figure}[t]
\centering
\includegraphics[width=\textwidth]{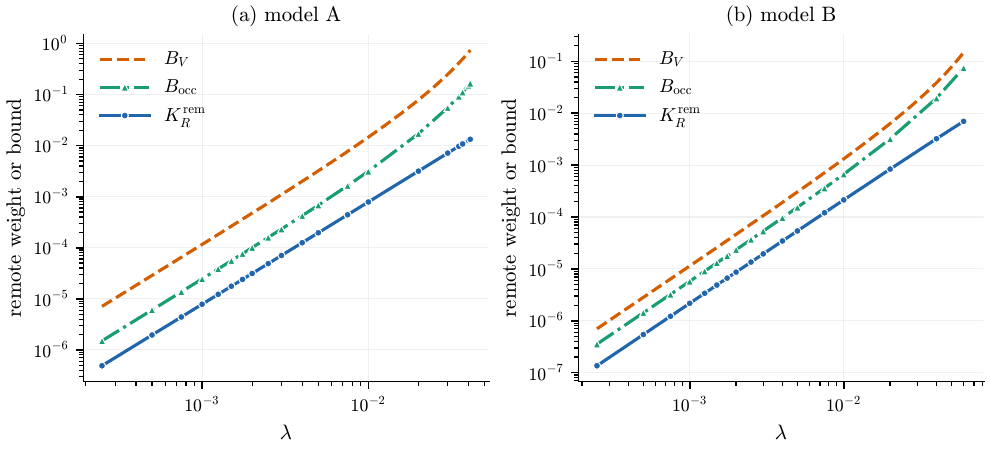}
\caption{Remote guiding-centre weight $K_R^{\rm rem}$ and the estimates $B_V$ of Eq.~\eqref{eq:gradientbound} and $B_{\rm occ}$ of Eq.~\eqref{eq:occupiedbound} for A and B.}
\label{fig:remainder}
\end{figure}

Figure~\ref{fig:remainder} compares the uniform estimate, the occupied-expectation refinement, and the actual remote weight. At $\lambda=0.01$, replacing $B_V$ by $B_{\rm occ}$ lowers the overestimate from factors 18.55 to 3.98 in A and 6.19 to 3.13 in B. The uniform curves in Fig.~\ref{fig:remainder} use the analytic constants of Table~\ref{tab:constants}. At the weakest amplitude the occupied-expectation estimate exceeds the remote weight by factors of only 3.07 in A and 2.62 in B. Retaining all three moments also avoids the cutoff surrogate. At $\lambda=0.01$, the three contributions to the two-moment slack in Eq.~\eqref{eq:slack} are $0.006935$, $0.013817$, and $0.003487$ in A, compared with $0.002676$, $0.001106$, and $0.079914$ in B. Thus improving the estimate matters most in A, while B is dominated by genuine geometric excess. Using the actual low-frequency upper edge gives two-moment left sides $0.502856$ and $0.413319$, much closer to the exact three-moment weights $0.502700$ and $0.413034$.

The finite-amplitude calculations use $n_{\max}=24$ and a $24\times24$ midpoint quadrature. Joint refinement to $n_{\max}=32$ and $32\times32$ at $\lambda=0.04$ in A and $0.06$ in B changes the reported dimensionless weights by less than $2\times10^{-9}$. The reconstructed dichroism plus the remote curvature agrees with $\kappa$ to $8\times10^{-11}$ over the dichroic scan through $\lambda=0.04$. Over the same scan, the three normalized moment combinations in Eq.~\eqref{eq:dichroism} reproduce the sum-rule values $t_r$, $2\nu$, and $\nu$ to within $2\times10^{-9}$, with the largest residual controlled by the Landau cutoff. These checks distinguish finite-basis accuracy from the much larger, physically necessary remote contribution.

For the higher-harmonic potential of Ref.~\cite{KIB2019}, at $\lambda=1.14\times10^{-4}$ the values are $K_{\rm sub}=0.50228$, $K_R^{\rm low}=0.50328$, and $K_R^{\rm rem}\simeq2.59\times10^{-5}$. Inserting spatially sampled gradient maxima into the uniform expression gives approximately $0.0455$, about $1.75\times10^3$ times the remote weight. This last number is an illustrative sampled estimate, not a certified supremum bound. It shows why occupied-expectation and spectral-separation refinements can be useful when many Fourier harmonics are present.

\FloatBarrier
\section{Physical scope and measurement considerations}
\label{sec:discussion}
Applying the finite-window bounds to a measured spectrum requires a physical regime in which the isolated-cluster and optical assumptions are simultaneously satisfied. The following discussion distinguishes possible implementations from the model established here and identifies the calibration requirements of the reconstruction.

\subsection{Physical regime and candidate implementations}
The optical bounds assume a clean, noninteracting parabolic band with a smooth scalar potential, a global insulating mini-gap, and temperature well below that gap. The a priori window further requires $\delta<\Ec/4$. These assumptions distinguish the present result from a general statement about moir\'e bands or lattice Hofstadter models. Electronic lateral superlattices realize magnetic subband physics~\cite{Albrecht2001}, but their interactions can invalidate the independent-particle regime. For illustration, take a square period of $100\,\mathrm{nm}$, flux $p/q=2$, and GaAs parameters $m=0.067m_e$ and $\epsilon_r=12.9$. Then $B\simeq0.827\,\mathrm T$, $\ellB\simeq28.2\,\mathrm{nm}$, and $\Ec\simeq1.43\,\mathrm{meV}$. For model A with $\operatorname{osc}V=0.5\,\mathrm{meV}$, $\delta/\Ec\simeq0.175$ and $\lambda\simeq0.029$. The projected global mini-gap is approximately $0.069\,\mathrm{meV}$ and the uniform estimate is $B_V\simeq0.23<1/2$. However, the unscreened Coulomb scale $e^2/(4\pi\epsilon_0\epsilon_r\ellB)\simeq4.0\,\mathrm{meV}$ exceeds both the cyclotron energy and the mini-gap. This example illustrates the optical scales, not an established noninteracting realization. Screening, disorder broadening, and temperature would all require separate control.

A possible continuum implementation is a rapidly rotating, spin-polarized Fermi gas with a weak co-rotating periodic potential. In the bulk, rotation supplies the quadratic magnetic Hamiltonian with $\omega_c=2\omega_{\rm rot}$ for rotation frequency $\omega_{\rm rot}$ after the residual trapping potential is compensated. Suppression of $s$-wave scattering between identical fermions favours an independent-particle regime, provided residual interactions and temperature remain below the mini-gap. Geometric squeezing has demonstrated access to Landau-level physics in rotating Bose condensates~\cite{Fletcher2021}, while circular-driving experiments establish the spectroscopy protocol in optical lattices~\cite{Tran2017,Asteria2019}. These are separate ingredients, not a demonstration of the complete fermionic continuum setting assumed here. For neutral particles, the measured excitation rate must be converted using the calibrated driving force in place of the electric coupling. The circular matrix elements and frequency reweighting are unchanged. Because $a_\pm(\omega)/\omega$ is proportional to the circular excitation rate $\Gamma_\pm(\omega)$, each moment $M_j$ corresponds to $\int\omega^{j+1}\Gamma_\pm\,\dd\omega$ and the weights $(1\mp\omega/\omega_c)^2$ multiply the rates directly. A flux $p/q$ per primitive cell requires a lattice period $a=\ellB\sqrt{2\pi p/q}$, which is $3.5\,\ellB$ for $p/q=2$. With magnetic lengths of a few micrometres this favours projected potentials over standard optical lattices.

\subsection{Calibration and scope of the optical test}
The reconstruction uses the bare kinetic parameter $\omega_c=|q_eB|/m$ in the chosen orientation and the absolute normalization of both circular conductivities. A periodic potential can shift the observed resonance, so its peak is not a substitute for this parameter. An unpatterned parabolic reference provides a possible calibration through the translationally invariant cyclotron mode protected under Kohn's hypotheses~\cite{Kohn1961}. For a fractional calibration error $\epsilon_c$ in $\omega_c$, differentiation of Eq.~\eqref{eq:dichroism} at fixed measured spectrum gives
\begin{equation}
 \delta D_R^{\rm low}
 =\frac{4\hbar}{e^2}
 \left(\frac{M_0^\Sigma}{\omega_c}-\frac{M_1^D}{\omega_c^2}\right)\epsilon_c
 +O(\epsilon_c^2).
 \label{eq:calibration}
\end{equation}
In model A at $\lambda=0.01$, a one-percent error shifts the reconstructed dichroism by approximately $1.9\times10^{-4}$, compared with its physical residual $1.1\times10^{-4}$. Resolving that residual requires a cyclotron-frequency error below roughly $0.6\%$. A common multiplicative error in the absorption normalization also multiplies $D_R^{\rm low}$, so the same residual would require normalization at about $0.02\%$ or better. These demands concern resolving the small residual, rather than testing the looser uniform bound. Differential channel errors, linewidths, and unresolved spectral tails require their own experimental error budget.

The noninteracting limitation concerns the curvature and gap-label argument, not the kinematic identity alone. Coordinate-dependent interactions between particles with a common charge-to-mass ratio commute with the total position. A uniform probe therefore retains the circular identity for total dipoles. Extending the finite-rank curvature argument to interacting or fractional states requires a many-body geometric construction and is outside the present result.

\section{Conclusions}
\label{sec:conclusions}
The projectively flat parent geometry fixes a gap-label-dependent lower bound on internal Landau-subband quantum weight, including for Chern-trivial occupied sectors. Its second fundamental form also identifies the circular transition weight measuring departure from saturation. At finite cyclotron gap, the integrated relative curvature persists, although uncorrected absorption does not measure the projected metric directly. Exact circular-dipole reconstruction instead recovers the guiding-centre weight and signed dichroism.

The finite-window bounds quantify missing remote transitions through the potential gradient and cyclotron separation, with quadratic suppression at weak coupling. Our two- and three-subband examples demonstrate both the improvement afforded by reconstruction and why the remote contribution cannot generally be neglected, even for a spectrally separated window. These results give the magnetic gap-label constraint a controlled optical interpretation within the stated noninteracting continuum setting. A natural next step is to determine whether an analogous relative-curvature construction extends to interacting or fractional quantum Hall states, where the finite-rank occupied-bundle description used here must be replaced by an appropriate many-body geometry.

The geometry of embedded subspaces also suggests a connection to quantum computing. In many qubit platforms the computational subspace is embedded in a larger multilevel spectrum and is addressed by drives that also couple it to leakage levels, a mixing problem routinely treated by Schrieffer--Wolff methods~\cite{Bravyi2011}. Quantum geometric tensors have been measured with a superconducting qubit~\cite{Tan2019}, and periodic driving extracts the quantum metric from excitation rates~\cite{OzawaGoldman2018}. The separation developed here between an embedded subspace, the fixed curvature of its parent and the leakage outside a finite window could therefore inform geometric benchmarks for programmable quantum simulators of Landau-level physics, although such applications remain prospective.

\section*{Acknowledgements}
\paragraph{Funding information}
This work was supported by the University of Turkish Aeronautical Association under Scientific Research Project (THKU-BAP) No.~2025-2/7.

\appendix
\section{Landau-level conventions and optical normalization}
\label{app:conventions}
With the definitions in Eqs.~\eqref{eq:scales} and \eqref{eq:coordinates},
\begin{equation}
 [\pi_x,\pi_y]=i\hbar q_eB,\quad
 [\eta_x,\eta_y]=i\ellB^2,\quad
 [R_x,R_y]=-i\ellB^2,\quad
 [R_i,\eta_j]=[R_i,\pi_j]=0.
 \label{eq:commutators}
\end{equation}
The guiding centre preserves the Landau index, while $\bm\eta$ connects $n$ only to $n\pm1$. The oscillator expectation values are
\begin{equation}
 \langle n|\eta_x^2|n\rangle=\langle n|\eta_y^2|n\rangle
  =\ellB^2(n+\tfrac12),\qquad
 \langle n|\eta_x\eta_y|n\rangle=\frac{i\ellB^2}{2}.
\end{equation}
Since the interlevel position matrix elements are those of $\bm\eta$, these identities give Eq.~\eqref{eq:parentQGT}, including its $+i\epsilon_{ij}$ sign. In the LLL, $\eta_+=(\eta_x-i\eta_y)/\sqrt2$ raises the oscillator and is the active channel. Reversing the sign of $q_eB$ interchanges the circular labels and reverses the oriented curvatures.

For a field $E\cos\omega t$, comparison of the golden-rule absorbed power with $\tfrac12\operatorname{Re}\sigma_{ii}E^2$ per area yields
\begin{equation}
 \int_0^\infty\frac{\operatorname{Re}\sigma_{ii}(\omega)}{\omega}\,\dd\omega
  =\frac{e^2}{2\hbar}K_{ii},\qquad
 K_{ij}=\frac{1}{2\pi}\int_{\MBZ}g_{ij}\,\dd^2k.
 \label{eq:normalization}
\end{equation}
For an occupied-to-empty derivative $d_i=\langle u_b|\partial_i u_a\rangle$, the circular strengths are
\begin{equation}
 \left|\frac{d_x\mp id_y}{\sqrt2}\right|^2
 =\frac{|d_x|^2+|d_y|^2\pm2\operatorname{Im}(d_x^*d_y)}{2}.
\end{equation}
The same formula applies with the parent covariant derivative for projected transitions and with $D$ for guiding-centre transitions. It fixes the signs in Eqs.~\eqref{eq:weak} and \eqref{eq:reconstruction}. In the strictly projected description the nondissipative guiding-centre drift contributes $\nu e^2/h$ to the Hall response. The intra-level spectral part supplies $(t_r-\nu)e^2/h$, giving the total $t_r e^2/h$.

\section{Curvature of the guiding-centre connection}
\label{app:curvature}
Work locally in the Bloch fibre with $u=e^{-i\bm k\cdot\bm r}\psi$. Then
\begin{equation}
 \bm\pi(\bm k)=\bm\pi+\hbar\bm k,\qquad
 \bm\eta(\bm k)=\bm\eta+\ellB^2\hat z\times\bm k,
 \qquad \partial_{k_x}\eta_y=-\partial_{k_y}\eta_x=\ellB^2.
\end{equation}
The interband position elements obey $\bm r_{ba}=i\langle u_b|\partial_{\bm k}u_a\rangle$. Subtracting $\bm\eta_{ba}$ yields $\bm R_{ba}=i\langle u_b|\bm D u_a\rangle$, which establishes the off-diagonal interpretation in Eq.~\eqref{eq:D}. Its curvature is
\begin{equation}
 [D_x,D_y]
 =i(\partial_{k_x}\eta_y-\partial_{k_y}\eta_x)-[\eta_x,\eta_y]
 =2i\ellB^2-i\ellB^2=i\ellB^2.
 \label{eq:centralD}
\end{equation}
The traced Gauss equation reads
\begin{align}
 \Tr\mathcal F^{PDP}_{xy}
 &=i\ellB^2r+
 \Tr\big[P(D_xP)(1-P)(D_yP)P-P(D_yP)(1-P)(D_xP)P\big]\nonumber\\
 &=i\ellB^2r+Q^R_{xy}-Q^R_{yx}.
 \label{eq:GaussD}
\end{align}
The difference between $PDP$ and the usual induced connection $P\dd P$ is the endomorphism-valued one-form $iP\eta P$. Tracing the curvature difference removes commutators, leaving
\begin{equation}
 \Tr\mathcal F^{PDP}=\Tr\mathcal F^{P\dd P}
       +i\dd\alpha,\qquad
 \alpha=\sum_i\Tr(P\eta_iP)\,\dd k_i.
 \label{eq:exactform}
\end{equation}
To see that $\alpha$ is global, let $\bm G$ be a reciprocal vector of the magnetic cell and use the Bloch sewing unitary $U_{\bm G}=e^{-i\bm G\cdot\bm r}$. On the fixed magnetic-periodic domain,
\begin{equation}
 P(\bm k+\bm G)=U_{\bm G}P(\bm k)U_{\bm G}^\dagger,
 \qquad
 \eta_i(\bm k+\bm G)=U_{\bm G}\eta_i(\bm k)U_{\bm G}^\dagger.
 \label{eq:sewing}
\end{equation}
The trace defining $\alpha$ is therefore compatible across opposite MBZ edges. It is also invariant under occupied-frame changes. Stokes' theorem on the torus gives $\int\dd\alpha=0$. Applying $(2\pi i)^{-1}\int$ to Eq.~\eqref{eq:GaussD} gives the central contribution $r\ellB^2|\MBZ|/(2\pi)=r/p$, while Eq.~\eqref{eq:exactform} identifies the total with $t_r$. This establishes Eq.~\eqref{eq:exactcurv} with the stated orientation.

The unbounded cyclotron coordinate requires a domain qualification. On a compact magnetic cell, the magnetic Schr\"odinger operators $H(\bm k)$ have a common second-order magnetic Sobolev domain and depend smoothly on $\bm k$. For smooth bounded $V$ and a uniformly isolated finite spectral cluster, the Riesz projector is smooth both in the Hilbert norm and into this operator domain by the fixed-domain resolvent construction~\cite{Kato}. Elliptic regularity places its finite-dimensional range in the domains of the first-order operators $\eta_i$ and their required derivatives. Thus $\eta_iP$, $D_iP$, and the finite-rank traces in Eqs.~\eqref{eq:GaussD}--\eqref{eq:exactform} are well defined and smooth. The commutators can first be computed on the smooth magnetic-periodic core and then extended to these ranges. The kinetic-energy estimate
\begin{equation}
 \|\bm\eta P\|_{\hs}^2
 =\frac{\ellB^4}{\hbar^2}\Tr(P\bm\pi^2P)
 \le \frac{2m\ellB^4}{\hbar^2}\,r(E_{\max}+\|V\|_\infty)
\end{equation}
provides a compatible finite bound. Smoothness into the operator domain, rather than this norm estimate alone, justifies differentiating the one-form $\alpha$.

\section{Perturbative separation of orbit and projector contributions}
\label{app:sw}
Let $P_n$ project onto free Landau level $n$, and write $P_0$ for the LLL. Following the block-diagonal perturbative construction~\cite{Bravyi2011}, choose an anti-Hermitian Schrieffer--Wolff generator $\mathcal S$ with
\begin{equation}
 P_n\mathcal SP_0=-\frac{\lambda P_nvP_0}{n\Ec},\qquad n\ge1,
 \qquad P_0\mathcal SP_0=0.
\end{equation}
The states $e^{\mathcal S}|\phi\rangle$ represent the low cluster, and block diagonalization yields Eq.~\eqref{eq:Heff}. Here $w$ has units of energy squared, so the correction $\lambda w/\Ec$ to the split-level operator has units of energy.

Using the oscillator matrix elements between $n=0$ and $n=1$ gives, for low-space interband matrix elements,
\begin{align}
 P_0e^{-\mathcal S}\bm r e^{\mathcal S}P_0
  &=\bm R-\frac{\ellB^2}{\Ec}P_0(\nabla V)P_0+O(\lambda^2),\nonumber\\
 P_0e^{-\mathcal S}\bm R e^{\mathcal S}P_0
  &=\bm R+O(\lambda^2).
 \label{eq:effectivecoordinates}
\end{align}
The remainder notation is an expansion in the dimensionless $\lambda$ at fixed $v$, $\Ec$, and $\ellB$. Its coefficients have the dimensions of the displayed coordinate. The second equation follows because $\bm R$ preserves the Landau index. The first equation agrees with expanding Eq.~\eqref{eq:dipole}, $r_\pm=(1\pm\omega_{ba}/\omega_c)R_\pm+O(\lambda^2)$, for low transitions.

For a positive-$\lambda$ projected insulator, let $\varepsilon_a(\bm k)$ be the eigenvalues of $\bar v$ and let $a$ and $b$ run over occupied and empty states within the LLL. Define
\begin{equation}
 \Delta^v_{ba}=\varepsilon_b-\varepsilon_a,\qquad
 d_i^{ba}=\langle u_b|\nabla_i^E u_a\rangle,\qquad
 F_{ab}=2\operatorname{Im}[(d_x^{ba})^*d_y^{ba}].
\end{equation}
The actual transition frequency is $\omega_{ba}=\lambda\Delta^v_{ba}/\hbar+O(\lambda^2)$. Expanding the circular strengths then gives the orbit contribution to the derivative of $K_{\rm sub}$,
\begin{equation}
 I_{\rm orbit}=\frac{1}{\pi\Ec}
 \int_{\MBZ}\sum_{a,b}\Delta^v_{ba}F_{ab}\,\dd^2k.
 \label{eq:orbitslope}
\end{equation}
The factor $1/\Ec$ is essential when units are restored. The projector contribution is
\begin{equation}
 II_{\rm proj}=\left.\frac{\dd}{\dd\lambda}
 K_{\rm gc}\!\left[\bar v-\frac{\lambda}{\Ec}w\right]\right|_{\lambda=0},
 \qquad
 \left.\frac{\dd K_{\rm sub}}{\dd\lambda}\right|_{0^+}
       =I_{\rm orbit}+II_{\rm proj}.
 \label{eq:projectorslope}
\end{equation}
The bracket denotes the internal quantum weight of the continued occupied projector of the indicated effective operator. For its derivative we use the four-point fourth-order stencil $[f(-2h)-8f(-h)+8f(h)-f(2h)]/(12h)$ with $h=10^{-5}$. It does not differentiate the overall energy scale of that operator. Equation~\eqref{eq:reconstruction} removes $I_{\rm orbit}$, so $\dd K_R^{\rm low}/\dd\lambda|_{0^+}=II_{\rm proj}$. The projector rotation leaves $\int F^{\rm in}=2\pi\kappa$ unchanged, although its metric and its local curvature distribution can both change.

\end{document}